\documentstyle[12pt,overcite,epsfig]{article}

\font\tenbf=cmbx10
\font\tenrm=cmr10
\font\tenit=cmti10
 1
\font\elevenrm=cmr10 scaled\magstep 1
 1

\newcommand{\labl}[1]{\label{#1}}
\catcode`\@=11
 
\def\({\relax\ifmmode[\else$[$\nobreak\hskip.3em\fi}
\def\){\relax\ifmmode]\else\nobreak\hskip.2em$]$\fi}
 
\def\gappr{\mathpalette\under@rel{>\approx}}
\def\lappr{\mathpalette\under@rel{<\approx}}
\def\gsim{\mathpalette\under@rel{>\sim}}
\def\lsim{\mathpalette\under@rel{<\sim}}
\def\under@rel#1#2{\under@@rel#1#2}
 
\def\under@@rel#1#2#3{\mathrel{\mathop{#1#2}\limits_{#1#3}}}
 
\def\under@@rel#1#2#3{\mathrel{\vcenter{\hbox{$%
  \lower3.8pt\hbox{$#1#2$}\atop{\raise1.8pt\hbox{$#1#3$}}%
  $}}}}

\def\parenbar{\mathpalette\p@renb@r}
\def\p@renb@r#1#2{\vbox{%
  \ifx#1\scriptscriptstyle \dimen@.7em\dimen@ii.2em\else
  \ifx#1\scriptstyle \dimen@.8em\dimen@ii.25em\else
  \dimen@1em\dimen@ii.4em\fi\fi \offinterlineskip
  \ialign{\hfill##\hfill\cr
    \vbox{\hrule width\dimen@ii}\cr
    \noalign{\vskip-.3ex}%
    \hbox to\dimen@{$\mathchar300\hfil\mathchar301$}\cr
    \noalign{\vskip-.3ex}%
    $#1#2$\cr}}}
 
\catcode`\@=12 % @ signs are no longer letters
\def\epem{e^+e^-}
\def\t12{\theta_{12}}
\def\theta{\vartheta}

\def\vth{\vartheta}

\def\phi{\varphi}

\def\ve{\varepsilon}

\def\rotwo{\rho^{(2)}}

\def\as{\alpha_s}

\renewenvironment{thebibliography}[1]
 { \elevenrm
   \begin{list}{\arabic{enumi}.}
    {\usecounter{enumi} \setlength{\parsep}{0pt}
     \setlength{\itemsep}{3pt} \settowidth{\labelwidth}{#1.}
     \sloppy
    }}{\end{list}}
 
\begin{document}

% title page for preprint
\mbox{ }\\
\hspace*{110mm} MPI-Ph/94-85\\
\hspace*{110mm} December 1994\\
\vfill
\begin{center}
{\bf QCD PREDICTIONS ON MULTIPARTICLE FINAL STATES}\footnote{\rm Invited       
    talk at the XXIV Int.\ Symp.\
    Multiparticle Dynamics, Vietri sul Mare, Italy, Sept.~1994,
    to be publ.\ in Proc., Eds. A. Giovannini, S. Lupia and R. Ugoccioni,
World Scientific, Singapore.}
\vfill
WOLFGANG OCHS\footnote{e-mail adress: wwo@dmumpiwh.bitnet}\\
\vspace{0.5cm}
{\sl
Max Planck Institut f\"ur Physik \\
Werner-Heisenberg-Institut\\
F\"ohringer Ring 6, D-80805 M\"unchen} \\
\vfill
{\bf Abstract}
\end{center}
\vspace{0.5cm}
\noindent We compare QCD predictions with experimental data on
inclusive multiparticle observables whereby we discuss various
realizations of the idea of parton hadron duality which is related
to a soft confinement mechanism. Special emphasis is given on
effects from the running $\alpha_s$, the soft gluon interference,
scaling properties and the role of the limiting scale $Q_0$ in
cascade evolution.
\vfill%
\mbox{ }
\newpage

%Here begins the paper for the proceedings
\begin{center}{{\tenbf QCD PREDICTIONS ON MULTIPARTICLE FINAL STATES\\}}
%              \footnote1Ù{to be published in
%Proc.\ of Cracow workshop on multiparticle production ``soft
%physics and fluctuations'', May 4-7 1993, Cracow, Poland, World
%Scientific Singapore.}
\vglue 2.0cm
{\tenrm WOLFGANG~OCHS\\}
\baselineskip=13pt
{\tenit Max-Planck-Institut f\"ur Physik\\}
\baselineskip=12pt
{\tenit F\"ohringer Ring 6, 80805 M\"unchen, Germany\\}
\vglue 0.8cm
\vglue 0.3cm
{\tenrm ABSTRACT}
\end{center}
\vglue 0.3cm
{\rightskip=3pc
 \leftskip=3pc
 \tenrm\baselineskip=12pt
\noindent
We compare
QCD predictions with experimental data on inclusive
multiparticle observables whereby
we discuss various realizations of the idea of parton hadron
duality which is related to a soft confinement mechanism.
Special emphasis is given on effects
from the running $\alpha_s$, the soft gluon interference,
scaling properties and the role of the limiting scale $Q_0$ in
cascade evolution.
\vglue 0.6cm}
\baselineskip=14pt
\noindent
{\bf 1. Introduction}
\par\vspace{0.4cm}
There is a hierarchy of Perturbative QCD predictions
depending on the sensitivity to the hadronization process
which is not yet understood at a satisfactory level within the
theory.\par
First there is the calculation of quantities in
Perturbation theory in fixed order of $\as$.
These studies strongly support Perturbative QCD as the theory of
strong interactions: The measurement of the total cross
sections in $\epem$ or in deep inelastic lepton nucleon
scattering, then the analysis of jet cross sections
in hard collisions. In the latter application a (not entirely trivial)
assumption has to be made on the equality of
cross sections for production
of hadron jets and of parton jets at the same resolution. The
spectacular successes of such predictions have established the basic
properties of the coupling and the vertices in PQCD at large
$Q^2$ in terms of one scale parameter $\Lambda_{\overline{MS}}$.\par
 
Secondly, by applying the leading log approximation for the
multiparton production,
one can push PQCD further and
resolve the more detailed intrinsic structure of jets.$^{1-6}$
An additional cutoff parameter $Q_0$ is
%\cite{GL,AP,KUV}. An additional cutoff parameter $Q_0$ is
introduced which regularizes the collinear and infrared singularities of
the gluon bremsstrahlung processes. Predictions can be obtained on
various observables of a multiparticle final state. There are
two different approaches towards comparison of such calculations
with experiment.
%\par\vspace{0.4cm}
%\noindent
 
{\sl 1. Parton-Hadron-Duality (PHD) in various forms}.\par
\noindent
The general idea is that a parton jet resembles in some aspects a
hadron jet. An initial argument for such a behaviour was the
proof of ``preconfinement'', i.e. the parton cascade already in the
perturbative region prepares color singlet clusters with finite
mass independent of energy \cite{AMATI}. However, these clusters are
too heavy for realistic phenomenological applications. So one has to
assume some kind of soft hadronization mechanism which allows to
compare properties of hadronic and partonic final states
\cite{BAS,DOK}. In the application one can distinguish further\\
a)  infrared and collinear safe observables\\
In this case the value of the observable doesn't change if a soft
particle is added or if one particle is
split into two collinear particles. Such observables are
independent of the cutoff $Q_0$ and have therefore a chance not to depend
on the final stage of the  jet evolution. Quantities of this type
are energy flows and -correlations, global quantities like thrust etc.
(for a review, see ref. 8).\\
%\cite{LEP}).\\
b) infrared sensitive observables\\
Here particles are counted. Examples are multiplicities, inclusive
spectra and correlations which are discussed in particular at a
multiparticle conference. These observables depend explicitly
on the cutoff $Q_0$ and the QCD results cannot be compared to data
immediately in a meaningful way.
 
For such quantities one may follow two strategies:\\
b1) One constructs again infrared safe quantities.
This is possible if the $Q_0$ dependence factorizes so that it dops out
after proper normalization. Another possibility is rescaling of
variables which may lead to asymptotically safe quantities.\\
b2) A more progressive strategy would be to interpret $Q_0$
as hadron mass (say $Q_0\sim m_\pi$) and to compare the observables
for a parton jet evolved down to hadronic scales
directly to the experimental data . This procedure has been shown to
work for momentum spectra (``Local Parton Hadron
Duality'' (LPHD) \cite{LPHD}).
%\par\vspace{0.4cm}
%\noindent
 
{\sl 2. Parton cascade with hadronization model.}\\
The parton cascade evolves down to a cutoff $Q_0>m_h$,
specific models are then introduced to describe the hadronization
process.
Most popular are the string model by the Lund group \cite{LUND}
($Q_0\sim 1$ GeV)
and the cluster model by Marchesini and Webber \cite{HERWIG}
($Q_0\sim 0.3$ GeV).
%\par\vspace{0.4cm}
 
In this review we discuss the infrared sensitive quantities in
multiparticle production. Also for such observables analytical
results are important and indispensable for a deeper understanding
of the QCD predictions; only then can  we
discuss power laws,
the consequences of running $\as$ and scaling properties.\par
 
The QCD calculations can be carried out analytically in the simplest
case in the double logarithmic approximation (DLA) which takes into
account the leading contributions from the infrared and collinear
divergencies and provides the asymptotic result at very high
energies. The nonleading corrections are typically large
and can be given by an expansion of the type
$1+a\sqrt{\as}+b\as +\ldots$
Such Modified LLA (MLLA) with subleading corrections
take into account, for example, energy recoil effects
but not yet angular recoils  (see, for example, ref. 6).%\cite{DOK}:
Alternatively one may derive QCD
results with Monte Carlo methods which are typically more
accurate and can serve as an important check of the approximations
of the analytical calculations.
 
The aim of the study of multiparticle phenomena in this context is
to learn about the soft limit of QCD and color confinement, the part
of the theory not well understood. In particular we are interested
how the running of $\as$, the soft gluon interference (``angular
ordering''), various scaling properties, the limiting scale $Q_0$
intrinsic to the parton cascade are reflected in the observable
particle distributions.
\begin{figure}
%\par\noindent
%\tenrm\baselineskip=12pt Fig.1
%\makebox[50mm]{\rule{0mm}{69mm}}
\centerline{\mbox{\epsfig{figure=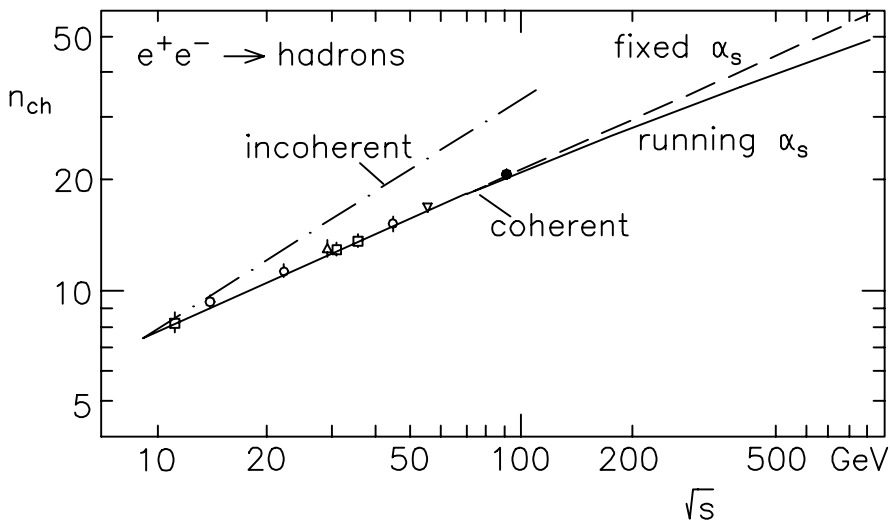,width=117mm}}}
\caption[...]{\protect\tenrm
Average charged particle multiplicity for different
CM energies $\sqrt{s}$ and NLO QCD
fit for running $\as$\cite{ALEPH}. Also shown is a fit with fixed $\as$
(power law) and the effect of
disregarding the soft gluon interference}
\end{figure}
\vspace{0.6cm}
\noindent
{\bf 2. Particle Multiplicities}\par
\noindent {\sl 2.1 Total Multiplicity}
%\par\vspace{0.4cm}
 
The multiplicity of partons emitted from a primary parton of momentum
$P$ into a cone of half opening angle $\Theta$
behaves asymptotically in QCD like
%\cite{NBMULT,QCDR}
\begin{equation}
\bar n\sim c\alpha^b_s
\left(\frac{P\Theta}{\Lambda}\right)^{2\gamma_0(P\Theta)}
\labl{nbar}
\end{equation}
For the full event one sets $\Theta=1$.
Here $\gamma_0=\sqrt{6\as/\pi}$ is the QCD anomalous
dimension controlling the multiplicity evolution in DLA\cite{MULT}.
For running
$\as$ we have $\gamma_0^2(p_T)=\beta^2/\ln(p_T/\Lambda)$ with
$\beta^2=12(\frac{11}{3} N_c-\frac{2}{3} N_f)^{-1}$.
The prefactor comes from the NLO of the anomalous dimension \cite{MULT1}
and $b=0.4916$ for $N_f=5$. For
fixed $\as$ the multiplicity behaves as in (\ref{nbar}) but with constant
exponent $\gamma_0$ and rises like a power with energy. For
running $\as$ the rise is more slowly.
The $Q_0$ dependence sits in the prefactor, therefore the ratio
$n(P)/n(P_0)$ is infrared safe. Results on $\bar n$ of higher
accuracy and for the full energy range are also available \cite{QCDR}.
 
A fit of type (\ref{nbar}) is shown in Fig.~1 \cite{ALEPH}. The distinction
between fixed and running $\as$ is not possible with present data
but the difference will become larger at higher energies.
The slope is well reproduced by
QCD\footnote{For $\Lambda =0.145$
corresponding to $\as (M_z)=0.117$; a change of $\Lambda$ by
a factor of 10 would change the slope by $\sim 15$\%.)}
%=======================ende der fussnote================================
with soft gluon interference taken into account. Neglecting
the interference would increase the slope by $\sqrt{2}$ clearly
inconsistent with data for any reasonable $\Lambda$ parameter.
\pagebreak
 
\par\vspace{0.4cm}
\noindent
{\sl 2.2 Multiplicity Distribution}
 
In the DLA, valid for asymptotic energies, one derives \cite{KNOQCD}
a scaling property for the probability $P_n$ in the rescaled
multiplicity $n/\bar n$ (``KNO-scaling'' \cite{KNO,POLYA}).
\begin{equation}
\bar n P_n = f(n/\bar n)
\labl{knof}
\end{equation}
There are large corrections to the asymptotic behavior from momentum
conservation alone which has been studied analytically
\cite{CT,DOKMU} and numerically \cite{DM}. In Fig.~2
we show as an example a calculation for a gluon jet with fixed
$\as$ in next-to-next-to leading order \cite{DOKMU}.
The asymptotic behavior for large $x$ is exponential
$f(x)\sim e^{-\beta_o x}$ (with $\beta_o\approx -2.552$).
The approach to this scaling limit is very slow and the preasymptotic
distribution at LEP energies looks quite different, at large
$x$ the distribution $f(x)$ drops faster, like
\begin{equation}
f(x)\sim exp (-\(Dx\)^\mu),\qquad \/\mu=(1-\gamma)^{-1}>1,\quad\/D\approx C
     (2\gamma/\pi)^\gamma    \labl{knodok}
\end{equation}
where the anomalous dimension
$\gamma \approx 0.41$ at LEP energies in this approximation and
$C\approx 2.5527$.
\begin{figure}
%\par\noindent
%\tenrm \baselineskip=12pt Fig.2
\makebox[50mm]{\rule{0mm}{95mm}}
\caption[..]{\protect\tenrm
The KNO multiplicity distribution in $x=n/\bar n$
for infinite energies (thin line) and LEP energies
(thick line), calculated for a gluon jet with fixed $\alpha_s$
in NNLO of QCD\cite{DOKMU}.
The negative binomial distribution (open points)
with parameter $K=7$ is also shown for comparison.
}
\end{figure}
%\par\vspace{0.4cm}
 
Analytic results for the realistic case of quark jets with running
$\alpha_s$ have been derived for moments of the multiplicity
distributions $R_q=<n(n-1)\ldots (n-q+1)>/<n>^2$. These normalized
quantities are again infrared safe. The QCD prediction on $R_2$
in comparison with experimental data are shown in Fig.~3. The DLA
result at infinite energies is $R_2=\frac{11}{8}$.
The next-to-leading order result\cite{MALW}
\begin{equation}
 R_2=\frac{11}{8}(1-\chi\sqrt{\as})\labl{rcor}
\end{equation}
with $\chi=0.55$ reduces the moment by $\sim 30$\% but still
differs from data by $\sim$~10\%. We have calculated this moment
also with the MC method (using the program HERWIG \cite{HERWIG}
at the parton level\footnote{We take the perturbative cascade
without $g\to q\bar q$ splitting at the end of the cascade and choose parameters
$\Lambda=0.15$ GeV, $m_q=m_g=0.32$ GeV.})
which fully takes into account energy momentum
conservation. This result finally matches the data and confirms PHD
for an infrared safe quantity.\par
\begin{figure}
%\par\noindent
%\tenrm \baselineskip=12pt Fig.3
\centerline{\mbox{\epsfig{figure=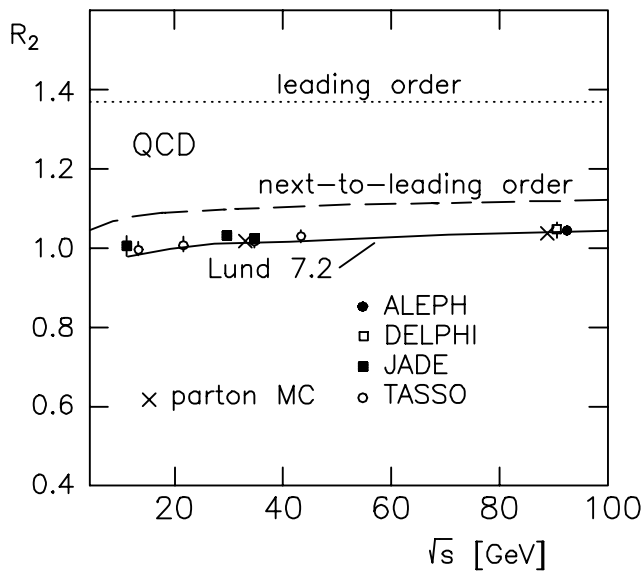,width=84mm}}}
%\makebox[50mm]{\rule{0mm}{77mm}}
\caption[..]{\protect\tenrm
Two particle multiplicity moment $R_2=<n(n-1)>/<n>^2$ vs.\ CM
energy. The QCD results in leading order (DLA), NLO \cite{MALW} and
from parton MC (HERWIG) in comparison with experimental data
\cite{ALEPH}.}
\end{figure}
\par
\vspace{0.6cm}
\noindent
{\bf 3. Inclusive One-Particle Distributions and the
          LPHD Hypothesis}\par
\noindent
{\sl 3.1 Momentum Spectra}
%\par\vspace{0.4cm}
 
One defines the logarithmic variable $\xi=\ln (P/k)\equiv \ln (1/x)$
for a particle of momentum $k$ emitted from a primary parton $P$;
$0\leq\xi\leq Y$ with $Y=\ln (P\Theta/Q_0)$. The asymptotic DLA
results are known for fixed $\alpha_s$ \cite{QCDR}
\begin{eqnarray}
\frac{dn}{d\xi}&\sim&\left(\frac{Y-\xi}{\xi}\right)^{1/2}I_1
\left(2\gamma_o \sqrt{\xi(Y-\xi)}\right)    \labl{xiex}\\
&\approx& \exp (-2\gamma_o(\xi -\frac{Y}{2})^2/Y) \labl{xiasy}
\end{eqnarray}
and for running $\as$ where one obtains in Gaussian approximation
\cite{MUEL,BAS}
\begin{equation}
\frac{dn}{d\xi}\sim \frac{\bar n}
{((Y+\lambda)^{3/2}-\lambda^{3/2})^{\frac{1}{2}}}
\exp\left(-\frac{3\beta(\xi-\frac{Y}{2})^2}
                        {(Y+\lambda)^{3/2}-\lambda^{3/2}}\right)
\labl{xirun}
\end{equation}
with $\lambda =\ln (Q_0/\Lambda)$, $\beta$ defined above.
The constant $\alpha_s$ result (\ref{xiasy}) can be recovered from this
by the formal limit $\beta,\lambda \to \infty,~ \beta/\sqrt{\lambda}=\gamma_0$
fixed.
This Gaussian shape, known as ``hump-backed plateau'', is characteristic
of the destructive soft gluon interference resulting in the suppression
of small momenta. The maximum occurs for $\xi^*=\frac{Y}{2}$.
The width behaves like
\begin{eqnarray}
\sigma^2\sim Y & \quad &{\rm for~} \as{\rm~ fixed} \labl{S2F}\\
\sigma^2\sim Y^{\frac{3}{2}} &\quad &{\rm for~} \as{\rm~running}
\labl{S2R}
\end{eqnarray}
So the running of $\as$ influcences the shape of the distribution.
As $Y$ depends on $Q_0$ this distribution is not infrared safe.
However one can consider a high energy limit using the rescaled variable
\cite{LPHD}
\begin{equation}
 \zeta=\frac{\xi}{(Y+\lambda)}\equiv \frac{\ln(P/k)}{\ln(P\Theta/\Lambda)}
\labl{zetav}
\end{equation}
In great analogy to the case of angular correlations \cite{OW1} to be
discussed below we can construct a quantity with a scaling limit and
obtain from (\ref{xirun})
\begin{equation}
\frac{\ln dn/d\xi}{\ln \bar n}\simeq 1-
\frac{3(\zeta-\zeta_0/2)^2}{2(1-(1-\zeta_0)^{\frac{3}{2}})}
\quad\to\quad 1-\frac{3}{2}(\zeta-\frac{1}{2})^2
\labl{zetasy}
\end{equation}
where the last limit holds for large $Y$, $\zeta_0=Y/(Y+\lambda)\to 1$.
So this quantity becomes infrared safe for high energies.
 
Results in higher orders have been obtained by the St.~Petersburg group
\cite{LPHD} (for a review of various approximations, see
ref. 6). %\cite{QCDR}).
Again the corrections to the DLA are sizable. For example, the position of the
maximum $\xi^*$ is shifted away by an amount
of $O(\sqrt{\alpha_s})$
\begin{equation}
\xi^* = Y(\frac{1}{2}+a\sqrt{\frac{\alpha_s(Y)}{32N_c\pi}}
-a^2 \frac{\alpha_s(Y)}{32N_c\pi} +\ldots),\qquad
a=\frac{11}{3}N_c+\frac{2N_f}{3N_c^2}
\labl{zetash}
\end{equation}
In the application to experiment
these authors proposed the hypothesis of ``Local Parton Hadron
Duality'' \cite{LPHD}
\begin{equation}
\frac{dn}{d\xi}\Big|_{hadron}=const \frac{dn}{d\xi}\Big|_{parton,Q_0=m_h}
\labl{lphde}
\end{equation}
a proportionality of the spectra of hadrons to the one of partons
for a cascade evolved down to a hadronic mass, $Q_0=m_h$,
which is taken typically as $m_\pi$. Of course, such a hypothesis
cannot be correct for all observables (for example, the mass spectrum
of two partons does not show a $\rho$-meson), but it is interesting
to explore its validity for sufficiently inclusive quantities.
The relation (\ref{lphde}) is tested in Fig.~4, where on the
r.h.s.\ the ``limiting spectrum'' in MLLA is inserted which is
obtained from the general formula by letting $Q_0=m_h\to\Lambda$.
For charged particle
spectra an effective hadron mass of $m_{h}=253$~MeV is taken.
It is remarkable how well the shape of the
distribution and its energy evolution is fitted by the theoretical
prediction in terms of 2 parameters $(m_h, const)$ for
not too small momenta $(k\gsim 400$~MeV).
A closer inspection of Fig.~4 also shows that the deviation from the DLA
prediction (\ref{xirun}) both in the position of the peak $\xi^*$ and
in the energy dependence are well supported \cite{QCDR}.
 
\begin{figure}
%\par\noindent
%\tenrm \baselineskip=12pt Fig. 4
\makebox[50mm]{\rule{0mm}{82mm}}
\caption[...]{\protect\tenrm
Distributions in $\xi =\ln 1/x_p$ of charged hadrons at different
energies compared with the analytical MLLA formula and a distorted
Gaussian for parton cascade evolved down to a mass of 250 MeV,
taken from Ref. \cite{OPAL}.}
\end{figure}
%\par\vspace{0.4cm}
Recently it has been questioned \cite{BCU}
as to what extent this result provides
evidence for the soft gluon interference of QCD.
A model has been considered (JETSET 7.3) with a parton cascade,
cut off at $Q_0\sim 1$~GeV, either including or not including the
soft gluon coherence, followed by string hadronization. The parameters
of the hadronization model are adjusted in both cases so as to
reproduce the main features of the data. In Fig.~5 one can see the
$\xi$-spectra of gluons, from the coherent calculation
with Gaussian shape and the non-coherent calculation with a higher particle
density towards larger $\xi$, i.e.\ smaller momenta. Both models
after hadronization yield the same spectra for
charged particles which are
also in good agreement with data on charged particles of Fig.~4.
Therefore within this hadronization model, as stated,\cite{BCU}
no evidence can be claimed from the measurements in favor of the soft
gluon interference effect. On the other hand, evolving the parton
cascade further according to QCD from $Q_0\sim 1$~GeV to
$Q_0=m_h$, only the gluon distribution for the coherent case
would approach the data (as is asserted by Fig.~4). This example
nicely demonstrates the predictive power of the LPHD hypothesis
and the similarity of parton and hadron spectra at comparable
scales. On the other hand hadronization models have considerable
flexibility to bring quite different theoretical schemes at the parton
level into agreement with the data. In particular, the LPHD relation
(\ref{lphde}) is a consequence of the string model only for a special
set of model parameters.\par
\begin{figure}
%\par\noindent
%\tenrm \baselineskip=12pt Fig.5
\makebox[50mm]{\rule{0mm}{85mm}}
\caption[..]{\protect\tenrm
Distributions in $\xi=\ln 1/x_p$ from the string model (JETSET 7.3)
at the parton level (``Gluons'') with and without the color
coherence of the QCD included, and for charged particles ($C^\pm$)
of the final hadronic state (from ref.24).}
\end{figure}
\par\vspace{0.4cm}
\noindent{\sl 3.2 Hadron mass effects}
 
%\par\vspace{0.4cm}
In a more speculative application of the LPHD hypothesis the
distribution of heavier particles of mass $M$ is given as in Eq.
(\ref{lphde}) with the cutoff $Q_0=M>\Lambda$,
whereas for pions $Q_0=m_\pi\approx\Lambda$. In the
DLA the peak position $\xi^* =\frac{Y}{2}$, so for a heavier
particle the peak is shifted by
 
\begin{equation}
\Delta\xi =-\frac{1}{2}\ln \frac{M}{m_\pi}
\label{delxi}
\end{equation}
with respect to the pion independent of energy. The numerical
studies of the MLLA equation \cite{DKT} conform this energy
independence and $\Delta\xi = f(M/\Lambda)$. In Fig. 6a
one can see that the data for $\pi^o,\eta$ are indeed separated by
an energy independent amount. The peak position $\xi^*$ at LEP-energies
for various hadrons are shown in Fig. 6b together with the
prediction from DLA, Eq.~(\ref{delxi}) and MLLA extracted
from the published results \cite{DKT}. The $\xi^*$ value drops when
going from $\pi$ to $K/\eta$ similarly to the MLLA prediction but
then saturates: the relation $Q_0=M$ does not work well for heavier
particles.\par
\begin{figure}
%\par\noindent
%\tenrm\baselineskip=12pt Fig. 6
\vspace{9mm}
\hfill{\mbox{\epsfig{figure=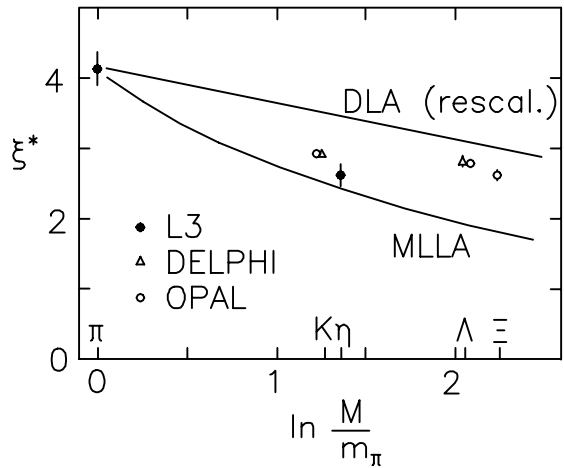,width=73mm}}}
%\makebox[50mm]{\rule{0mm}{70mm}}
\caption[...]{\protect\tenrm
Peak position $\xi^*$ of the distribution in $\xi =\ln(P/k) $
vs. CM energy for $\pi^0$ and $\eta$ \cite{PIETA} (left side);
$\xi^*$ vs.\ hadron mass $M$
at LEP energies compared with predictions from DLA and MLLA. Data
from compilation \cite{FOC} (right side).}
\end{figure}
\par\vspace{0.6cm}
\noindent
{\bf 4. Two Particle Correlations}\par
\noindent{\sl 4.1 Momentum Correlations}
%\par\vspace{0.4cm}
 
The normalized correlation function is defined by
\begin{equation}
R_2(\xi_1,\xi_2)=\rho^{(2)}(\xi_1,\xi_2)/\rho^{(1)}(\xi_1)\rho^{(1)}
(\xi_2)
\label{rmom}
\end{equation}
in terms of the $n$-particle density $\rho^{(n)}$. It has been calculated
in NLO as an expansion in the arguments up to second order \cite{FW}
and is found to depend only on the rescaled observables
$\zeta_i=\xi_i/(Y+\lambda)$ at high energies. Alternatively one could choose
the rescaled observable $\xi/\xi^*$ as $\xi^*=Y/2$ in DLA.
In Fig.~7 we show $R_2(\xi_1,\xi_2)$ for $\xi_1=\xi_2$ vs. $\xi/\xi^*$
(where $\xi\equiv \xi_1$) at the parton and hadron level as we obtained
from the HERWIG MC
at two different
primary quark energies.
 
Contrary to the integrated moment $F_2$ (see Fig.~3) PHD does not
work well for the differential moment but it improves with increasing
energy, so one may speak of asymptotic PHD.
\begin{figure}
%\makebox[50mm]{\rule{0mm}{65mm}}
%\par\noindent
%\tenrm \baselineskip=12pt Fig. 7
\centerline{\mbox{\epsfig{figure=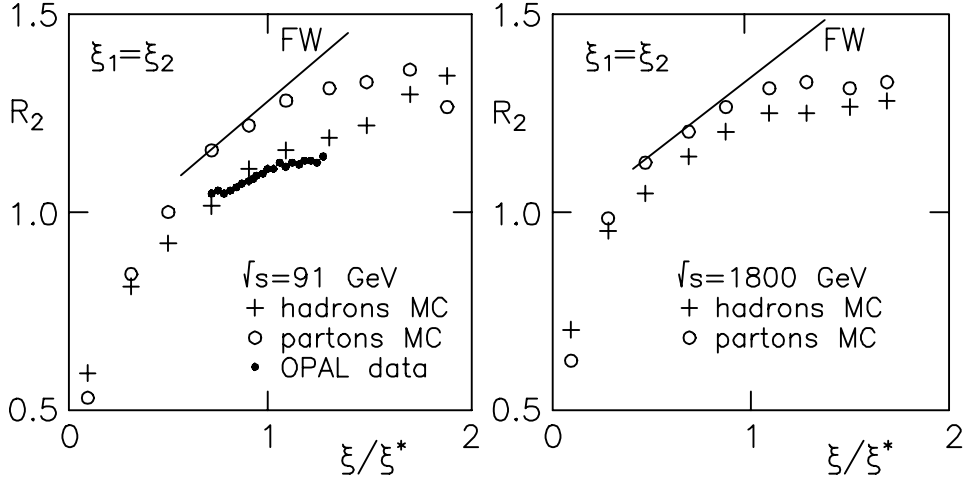,width=127mm}}}
\caption[...]{\protect\tenrm
Two-particle momentum correlation function $R_2(\xi_1,\xi_2)$ for
$\xi_1=\xi_2$ vs.\ $\xi/\xi^*$, where $\xi\equiv\xi_1$, and
$\xi^*$ is the peak
position of $\rho^{(1)}$, for hadrons and partons from the HERWIG
MC together with
the analytical result in linear approximation
by Fong and Webber ($\Lambda$=255~MeV).
Also shown are the OPAL data \cite{OPALM}.}
\end{figure}
\par\vspace{0.4cm}
\noindent
{\sl 4.2 Azimuthal angle correlations}
%\par\vspace{0.4cm}
 
Energy-multiplicity-multiplicity correlations in the aximuthal
angle around the jet direction have been calculated in DLA
\cite{DKMW} and next-to-leading order \cite{DMO}. The NLO
calculations for large relative angles $(\phi=\pi)$ differ from
the MC results at the parton level
by $\sim$~10\% but the experimental data follow
the MC results closely for not too small
relative angles $\phi$ \cite{OPALPH}
in favor of PHD. The difference between coherent and incoherent models
is rather small, typically about 5\%.
\par\vspace{0.4cm}
\noindent
{\sl 4.3 Polar angle correlations}
%\par\vspace{0.4cm}
 
The correlations in the relative polar
angle $\theta_{12}$ of two
partons within the forward cone of half opening angle $\Theta$
have been derived in DLA \cite{OW,OW1}.
This is a special case of $n$-particle angular observables
$h^{(n)}(\delta,\vartheta,P)$, like
multiplicity moments, to be discussed below. For such quantities
the leading asymptotic behavior is given by
\begin{eqnarray}
h^{(n)}(\delta,\vth, P) &\sim&\exp(2\beta\sqrt{\ln (P\vth/\Lambda})
  \omega(\ve, n)) \labl{hn}\\
\ve &=&\frac{\ln (\vth/\delta)}{\ln (P\vth/\Lambda)}   \labl{eps}
\end{eqnarray}
Because of $\delta\gsim Q_0/P\gsim\Lambda/P$
the rescaled angular variable fulfils $0\leq\ve\leq1$.
The scaling function $\omega(\ve,n)$ is known \cite{OW}
and can be
expanded for small $\ve$ like
\begin{equation}
\omega (\ve, n) = n-\frac{1}{2}\frac{n^2-1}{n}\ve +\ldots
\labl{omsm}
\end{equation}
and for large $n$ like
\begin{equation}
\omega(\ve, n)=n\sqrt{1-\ve}+O(\frac{1}{n})
\labl{omla}
\end{equation}
For the 2-particle correlation within the cone of half opening
$\Theta$ the quantity
\begin{equation}
\hat r(\vth_{12},\Theta,P)=\rho^{(2)}(\vth_{12},\Theta,P)/\bar n^2
(\Theta,P)
\labl{rhat}
\end{equation}
is considered where $\bar n$ is the multiplicity in the forward cone. In DLA
one obtains the asymptotic prediction
\begin{equation}
\hat r(\vth_{12},\Theta,P)\sim \exp\left(
2\beta\sqrt{\ln (P \Theta/\Lambda})
(\omega (\ve, 2)-2)\right)
\labl{rhas}
\end{equation}
An interesting property of this result is the ``$\ve$-scaling''.
Up to a known prefactor in the exponent the correlation function
depends on the three variables $\vth_{12},\Theta,P$ only
through the single rescaled angular variable $\ve$.
This scaling property is checked by the MC calculation at the
parton level in Fig.~8 which shows the $\ve$-dependent
part of the exponent in (\ref{rhas}) vs.\ $\ve$. For
sufficiently small $\ve\lsim 0.5$ this scaling property
is indeed well satisfied for $P\gsim 20$~GeV.
Eq. (\ref{rhas}) can also be seen to reproduce the trend of the
MC data. Some nonasymptotic corrections within DLA different for
quark and gluon jets, are also known\cite{OW1}. In Fig.~8 we have adjusted
the overall normalization because this is a nonleading effect.\par
\begin{figure}
%\makebox[50mm]{\rule{0mm}{135mm}}
%\par\noindent
%\tenrm \baselineskip=12pt Fig. 8
\centerline{\mbox{\epsfig{figure=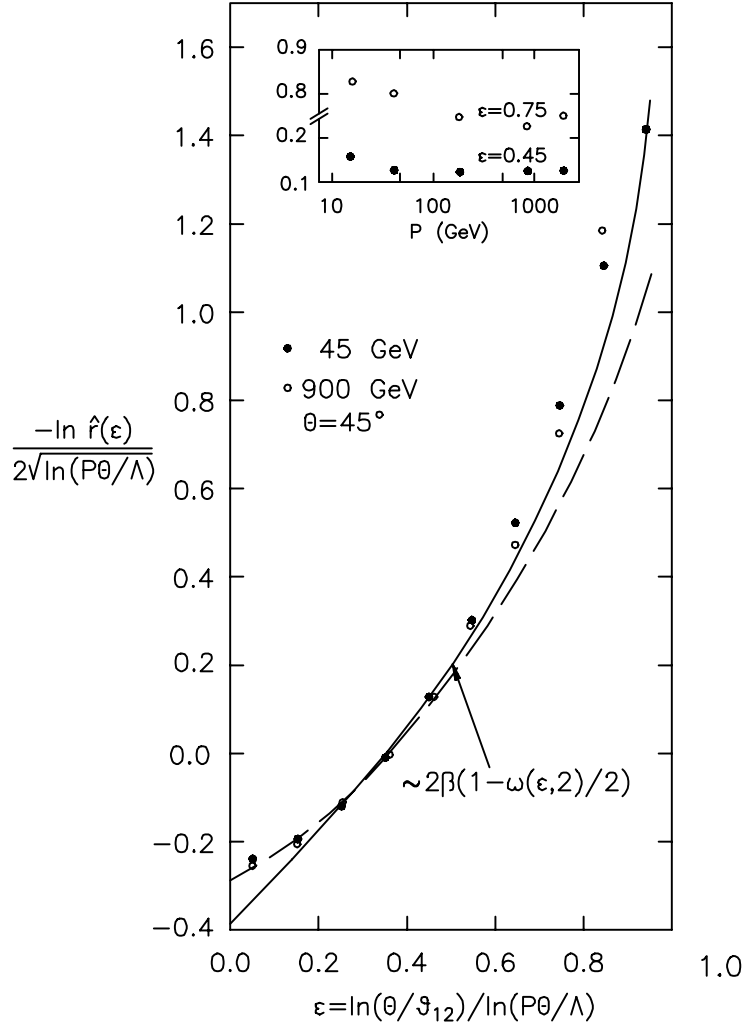,width=97mm}}}
\caption[...]{\protect\tenrm
Rescaled 2-particle polar angle correlation vs.\ scaling variable
$\ve$ for different primary energies $P$ and for fixed cone opening $\Theta$
(with respect to the sphericity axis)
from the parton MC
(HERWIG). The full curve represents the high energy limit of the
DLA; the dashed curve the prediction for quark jets at 45 GeV; the
normalization of the curves is adjusted. The insert shows the
energy dependence of the same quantity for fixed $\ve$; from
Ref.\cite{OW1}.}
\end{figure}
%\par\vspace{0.4cm}
For small $\ve$ with (\ref{omsm}) the correlations become
asymptotically power behaved
\begin{equation}
\hat r(\vth_{12})=
(\Theta/\vth_{12})^{-\frac{3}{2}\gamma_o(P\Theta)}\labl{rhatpow}
\end{equation}
and this can be related to the
selfsimilarity of the jet cascade. This power law holds asymptotically
in the full angular region for fixed $\alpha_s$, so the
curvature in Fig.~8 in the asymptotic curve reflects the
running of $\as$.
 
A comparison of the parton and hadron MC reveals that in the
region with scaling $\ve\lsim 0.5$ there is also PHD
(in the MC). This region is again characterised by its
independence of $Q_0$ in confirmation of the general rule.
A preliminary result from the DELPHI collaboration \cite{DELPHI},
also presented at this conference, supports indeed PHD for this
observable, i.e.\ the close similarity of the experimental and parton
MC data
(as $Q_0>m_h$ in the MC we talk here about PHD and not LPHD).
Furthermore the first evidence for $\ve$-scaling in the
opening angle $\Theta$ (for $30^o\leq\Theta\leq 60^o$) has been
presented.
 
Another interesting aspect is the sensitivity to the soft gluon
interference which is most naturally observed in the polar angle
correlations. This is demonstrated by an analysis with L3 data
\cite{SYED} considering the
Particle-Particle-Correlations (PPC). They are defined like the
well-known energy-energy correlations \cite{EECOR} but without
the energy weights in the double sum over particle pairs. Likewise
defined is
the asymmetry PPCA
$(\vth_{12})$ = PPC $(180^o-\vth_{12})$ - PPC($\vth_{12})$.
Disregarding the fluctuations of multiplicity $N_{ch}$
this quantity is related to our correlation function like
PPCA$(\vth_{12})\approx -(\rotwo
(\vth_{12})-\rotwo_{{\rm uncorr.}}(\vth_{12}))/\bar n^2$.\par
\begin{figure}
%\makebox[50mm]{\rule{0mm}{85mm}}
%\par\noindent
%\tenrm \baselineskip=12pt Fig. 9
\centerline{\mbox{\epsfig{figure=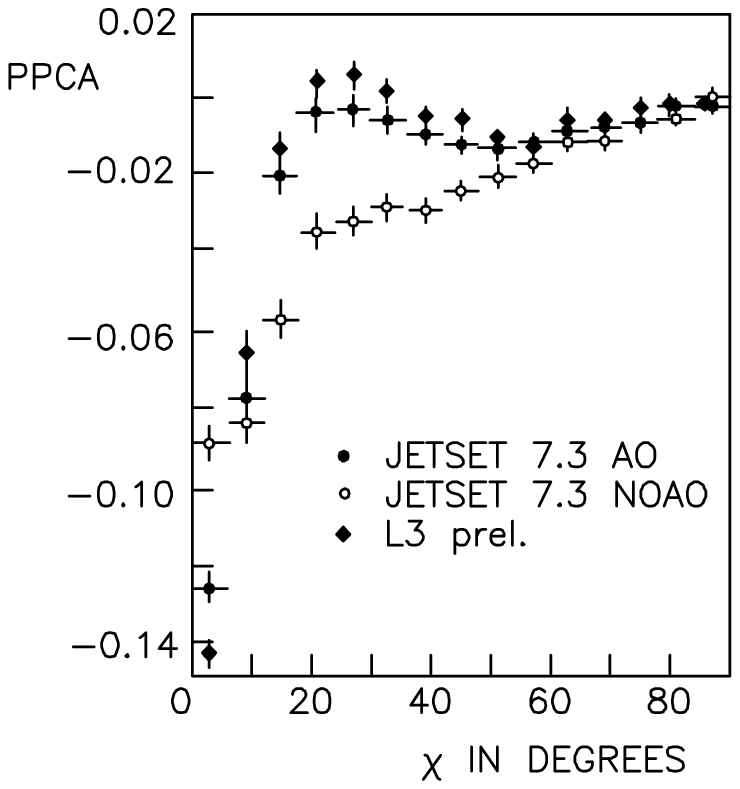,width=75mm}}}
\caption[...]{\protect\tenrm
Particle Particle Correlation Asymmetry vs. relative polar angle
$\chi\equiv \theta_{12}$ from L3 experiment
\cite{SYED} in comparison with the string MC with angular ordering
(AO) and without (NOAO).}
\end{figure}
%\par\vspace{0.4cm}
 
In Fig.~9 results are shown
from the JETSET MC for the angular asymmetry PPCA which varies strongly
for small $\vth_{12}$ if angular ordering (AO) is properly
included as required by QCD. The data follow closely the AO-case.
The sharp dip at small $\vth_{12}$ is related to the peak in
$\rotwo (\vth_{12})\sim 1/\vth_{12}$ in DLA \cite{OW},
but also Bose-Einstein correlations could be important in this region.
 
The angular ordering condition implies that a soft gluon is
emitted from a parton 1 only within an angular cone limited by
the next color connected parton ($\vth_{12}\leq\vth_{{\rm 1~next}}$).
We may estimate roughly $\vth_{{1~next}}\sim\bar\vth_{12}\sim\Theta/
\sqrt{\bar n}$ which corresponds to about
$1/\sqrt{15}{~\rm rad}\sim 15^o$.
One expects the emission within this angular region to be enhanced,
outside to be suppressed in comparison to the not ordered case.
Interestingly, such an effect is indeed seen in Fig.~9 (note that
PPCA $\sim -\rotwo$), so this interference effect between
gluons is actually visible between hadrons in the region 15-50$^o$.
This consideration also makes clear that the effect of angular
ordering is most clearly seen in the polar angle (and not in the
azimuthal angle or between momenta).
\pagebreak
\par\vspace{0.6cm}
\noindent
{\bf 5. Angular Correlations of General Order $n$}
\par\vspace{0.4cm}
The general $n$-particle cumulant correlation function has also
been studied in DLA \cite{OW1}. Here we discuss only as
application the integral over the correlation function in certain
angular regions. First we discuss the sidewise ring where the
polar angles $\vth_i$ with respect to the initial parton are
within the range $\vth -\delta\leq\vth_i\leq\vth +\delta$,
second the sidewise cone centered at a polar angle $\vth$ and
half opening $\delta$. We refer to these configurations as
dimension $D=1$ and $D=2$ cases according to the phase space
volume $\delta^D$.
The factorial multiplicity moments are obtained from
\begin{equation}
f^{(n)}(\vth,\delta)=\int\rho^{(n)}(\Omega_1,\ldots\Omega_n)
d\Omega_1\ldots d\Omega_n
\labl{fmom}
\end{equation}
or normalized $F^{(n)}=f^{(n)}/\bar n^n$. In analogy
the cumulant moments are constructed from the cumulant
(connected) correlation functions and are related to the
factorial moments\cite{INTSUM} (i.e.\ $C_2=F_2-1$, etc.).
 
At high energies one obtains in DLA \cite{OW,DD,BMP}
\begin{equation}
C^{(n)}(\vth,\delta)=\left(\vth/\delta\right)^{\phi_n}\/,
\/\varphi_n =D(n-1)-2\gamma_0 (P\vth)(n-\omega(\ve,n))/\ve
\labl{cnasy}
\end{equation}
In leading approximation the difference between factorial and
cumulant moments vanish.\footnote{The formula (\ref{cnasy})
is given for the cumulant moments in ref.~\cite{OW} and for
the factorial moments in refs.~\cite{DD,BMP}.}
%===================ende der fussnote==========================
Corrections in the MLLA are also given \cite{DD} and amount to
typically 10\%.
 
In the limit of small $\ve$ (sufficiently large opening angles
$\delta$) one can use the linear approximation for
$\omega(\ve, n)$ (\ref{omsm}) and (\ref{cnasy}) becomes a power
law with
\begin{equation}
\varphi_n=D(n-1)-(n-\frac{1}{n})\gamma_0(P\vth)
\labl{intm}
\end{equation}
This power behaviour reflects the fractal structure of the selfsimilar
cascade. In this limit, the behavior of moments is also independent
of $Q_0$, i.e.\ infrared safe. For small angles $\delta\sim Q_0/P$
the above asymptotic formula is inappropriate and strong
sensitivity to $Q_0$ appears.
 
This type of power behavior was studied intensively in the last
years in the context of ``intermittency'' \cite{BP}. Whereas
these phenomenological studies concentrate mainly on the small angle
region the power behavior in QCD occurs only for the fully
developed cascade at sufficiently large angles (small $\ve$).
 
As in the case of two particle correlations we can exhibit
the $\ve$-scaling by considering the quantity
\begin{equation}
-\hat C^{(n)}= -\frac{\ln\((\delta/\vth)^{D(n-1)}C^{(n)}\)}
                     {n\sqrt{\ln (P\vth/\Lambda)}}
\labl{chat}
\end{equation}
in the high energy limit. One obtains
\begin{eqnarray}
-\hat C^{(n)} &\sim& 2\beta (1-\omega (\ve,n)/n)\labl{chatlim}\\
              &\approx &2\beta (1-\sqrt{1-\ve}) \labl{chatap}
\end{eqnarray}
where the last approximation follows for the large $n$ limit
(\ref{omla}) and is independent of $n$.
 
In Fig.~10 we show the quantity $-\hat C^{(2)}$ for the ring
$(D=1)$ from (\ref{chat}) for different energies from the parton MC.
There is still considerable scale breaking at small $\ve$ but
the data points approach at high energies the DLA result
(\ref{chatlim}). We also show in Fig.~10 the analogous results
for factorial moments (definition as in (\ref{chat})). In this
case the $\ve$-scaling sets is already at low energies with an
$\ve$-dependence as in (\ref{chatlim}). Note that the absolute scale
as well as the difference between factorial and cumulant moments
is of nonleading order in DLA.
 
An interesting aspect of these results is the universal behavior
of the observables $\hat C^{(n)}$ for different $n,D$ and also of
the quite different observable $\hat r(\ve)$.
The observables $\hat r(\ve)$ and  $\hat C^{(2)}$ in Figs. 8 and 10
are determined from different parts of
phase space but nevertheless approach the same asymptotic limit.
Such a similarity is a characteristic property of the QCD cascade.
The comparison of parton and hadron MC shows again that PHD
works well in the model for not too large $\ve$ near 1 \cite{OW1}.
\begin{figure}
%\makebox[50mm]{\rule{0mm}{107mm}}
%\par\noindent
%\tenrm \baselineskip=12pt Fig. 10
\vspace{9mm}
\centerline{\mbox{\epsfig{figure=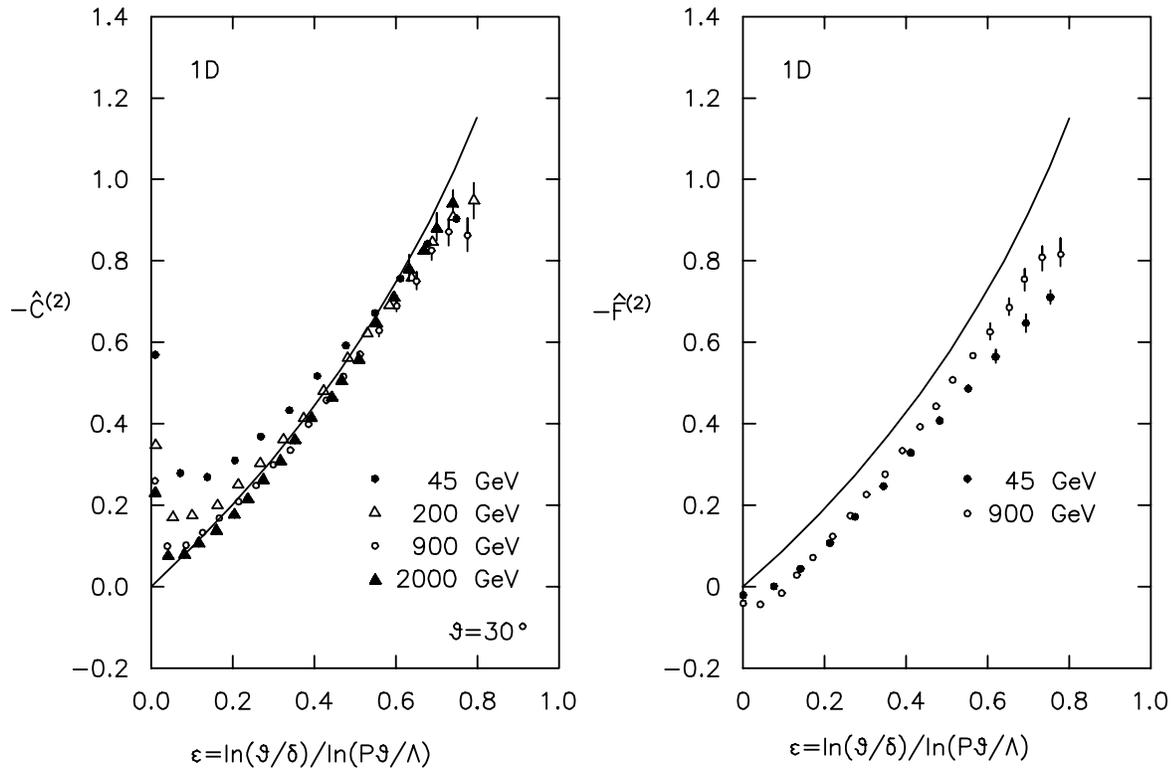,width=154mm}}}
\caption[...]{\protect\tenrm
(a) Rescaled cumulant moments for the ring (D=1) as defined in
Eq.(\ref{chat}) from the parton MC for different jet momenta $P$ in comparison
with the asymptotic prediction Eq.(\ref{chatlim}), (b) same as (a) but for
factorial moments.}
\end{figure}
\par\vspace{0.6cm}
\noindent
{\bf 6. Clans}
\par\vspace{0.4cm}
Clans are defined \cite{GVH} as group of particles of common ancestor.
The clans are independently produced and therefore the number of clans in
an event follow a Poisson distribution. With the assumption of a
logarithmic distribution of particles in a clan one obtains a
negative binomial distribution of particles, which describes the
data in a good approximation.
 
The analysis of multiplicity distributions in symmetric rapidity
windows $|y|<y_c$ has revealed the interesting result that the
average number of clans $\bar N(y_c,\sqrt{s})$ at fixed CM energy
$\sqrt{s}$ grows linearly with $y_c$ for a large range of $y_c$
before it bends if $y_c$ approaches the full phase space
$y_{fps}$. There is only a weak energy dependence of this phenomenon.
 
One may think of the clans as the jets which evolve from the gluons
radiated off the primary partons in the collision in a Bremsstrahlung
fashion with a Poisson distribution. This connection between
parton and hadron level has been studied in hadronization
models. In particular, the $1/k$ parameter of the negative
binomial distribution was found about equal at parton and hadron
level for
$\sqrt{s}\geq 200$~GeV whereas the multiplicity $\bar n$ was
larger for hadrons than for partons. It is interesting to note
that $1/k$ parameter determines fully the normalized factorial
moments. They are expected to be infrared safe as it is known for the
full interval (see 2.2 and Fig. 3).
On the other hand $\bar n$ is clearly
$Q_0$ dependent and therefore different at parton and hadron level.
In a generalization of these findings it has been suggested that
the differential rapidity distributions $\rho^{(n)} (y_1\ldots y_n)$
at the parton and hadron level are proportional (``Generalized Local
Parton Hadron Duality'' \cite{GLPHD}). This relation certainly works
best if the cutoff $Q_0$ is close to the hadron mass. Otherwise there
will be different kinematic limits. In that case there may be a
better duality for the rescaled rapidities $y/y_{max}$.
 
An interesting scaling property in terms of a rescaled rapidity
has recently been found \cite{UGL} for the clan multiplicity.
The multiplicity ratio $\pi^*=\bar N(y_c,\sqrt{s})/\bar N (y_{fps},
\sqrt{s})$ is calculated analytically in a simplified version of
the QCD parton shower. This quantity scales well in the rescaled
rapidity $y_c^*=y_c/y_{fps}$
\begin{equation}
\pi^*(y_c^*,\sqrt{s})\approx\pi^*(y_c^*)   \labl{pist}
\end{equation}
in the studied range $50<\sqrt{s}<500$~GeV and approaches in the
very high energy limit
\begin{equation}
\pi^*(y^*_c,\sqrt{s})\simeq y_c^*+O\left(\frac{1}{\ln\ln\sqrt{s}}\right)
\labl{pistas}
\end{equation}
where the approach to this limit is very slow.
\par\vspace{0.6cm}
\noindent
{\bf 7. Limitations of Parton Hadron Duality}
\par\vspace{0.4cm}
For the inclusive observables discussed so far the PHD
concept seems to work rather well as seen from experimental data
or suggested by MC calculations. We have, finally, to discuss
where we expect or observe the limitations.
 
We clearly expect deviations in the short range correlations due to
resonance effects which occur at the hadron but not at the parton
level, i.e.\ for masses $M_{ij}\lsim 1.-1.5$~GeV. This
causes a violation of PHD, for example, in the correlation function
$r(\ve)$ for $\ve\to 1$ or $\vth_{12}\to Q_0/P$. A recent example
of this type has been presented by the OPAL collaboration
\cite{OPALSUB}. The measured ratio of sub-jet multiplicities
in two- and three-jet events deviates from the QCD prediction \cite{CDFW}
at the parton level below a resolution scale of $\sim 2$~GeV.
As this scale is a bit large
it would be desirable to understand better the reasons for the
sudden change in slope at the hadron level (charm production?).
In any case the discrepency
between data and theory is only of the order of 10\%.
\footnote{I would like to thank T. Sj\"ostrand
for bringing this result to my attention and for correspondence.}
%================ende der fussnote================================
 
Another interesting limit wher PHD may fail and is expected to fail
in standard hadronization models is the quasielastic limit
in $\epem$ annihilation with a large
rapidity gap. Such events at the parton level correspond
essentially to $\epem\to q\bar q$+(few soft gluons). The
probability for this state is given by the Sudakov form factor
for no radiation into this angular interval. With
PHD the parton final state would transform into a similar final
state with large rapidity gap (see Fig.~11).\par
\begin{figure}
%\makebox[50mm]{\rule{0mm}{35mm}}
%\par\noindent
%\tenrm \baselineskip=12pt Fig. 11
\centerline{\mbox{\epsfig{figure=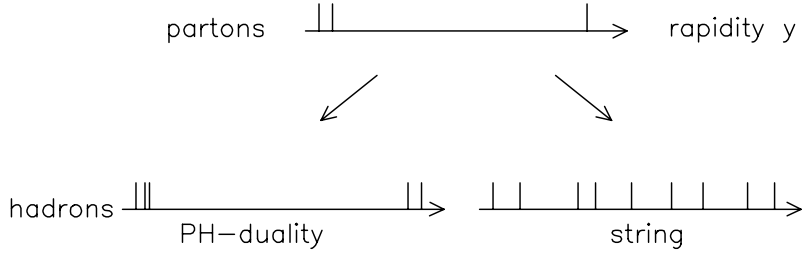,width=105mm}}}
\caption[...]{\protect\tenrm
Hadronization of a quasi-exclusive partonic state
yields a different hadronic final state according to Parton
Hadron Duality and the string model.}
\end{figure}
%\par\vspace{0.4cm}
 
In the
standard hadronization models there is a color field between the
separated partons which would decay into many hadrons of small
$p_T$.
Assuming an independent emission of particles or clusters with
rapidity density $c$ the probability for a gap of length
$\Delta y$ is \cite{LST}
$P=\exp(-c\Delta y)$, so there is this additional
suppression to obtain events with a large rapidity gap.
Then there is no PHD
in the standard hadronization models in this limit.
If an effect of this type was observed it would imply color
bleaching earlier in the cascade than usually thought, some kind
of exclusive PHD \cite{CTO}.
%\par\vspace{0.4cm}
 
An effect of this type has actually been
seen in the MC simulation of the 2-jet rate for small
resolution scales $y_{{\rm cut}}$ at parton and hadron level\cite{WEBAA}.
For $y_{{\rm cut}}\sim
10^{-4}$ ($K_T$ scale $\sim$1~GeV) the rate for the hadronic
2-jet events is found about hundred times larger than for the partonic ones.
\par\vspace{0.6cm}
\noindent
{\bf 8. Conclusions}
\par\vspace{0.4cm}
It is quite remarkable that the QCD calculations on the parton
cascade match the experimental data and this suggests indeed a
rather soft hadronization mechanisme. The scheme
of PHD is well-defined and economic with the only parameter
$\Lambda$ for infrared safe quantities, and the additional
quantity $Q_0$ otherwise. We have considered here mainly
inclusive infrared sensitive quantities.\par
\par\vspace{0.4cm}
\noindent
{\sl 1. PHD and LPHD}.
As a general rule, if the DLA result is independent of $Q_0$ (after
appropriate normalization) then the hadronization corrections are
small (multiplicity moments, angular observables for small $\ve$).
There are cases where the $Q_0$ dependence disappears only
asymptotically, then also PHD does the same (momentum
correlations).
In some cases the nonasymptotic corrections are large (cumulant
angular correlations $\hat C(\ve)$).
The LPHD hypothesis with $Q_0=m_h$ is
successful for momentum spectra of the light hadrons (not well
for heavier hadrons), but no other
application has been provided so far.
\par\vspace{0.4cm}
\noindent
{\sl 2. Rescaled variables} appear naturally in the description
of the QCD cascade: $n/\bar n,\zeta,$ $\ve,$$y^*_c$. Especially the
recently introduced angular variable $\ve(\delta,\vth,P)$
provides a new type of
scaling predictions for the QCD cascade
with two redundant variables $\vth,P$
\cite{OW,OW1}.
First results presented at this meeting look promising for the
$\vth$-independence \cite{DELPHI}.
\par\vspace{0.4cm}
\noindent
{\sl 3. Soft gluon interference} is taken into account by the angular
ordering prescription. There is evidence from the shape of the
hump backed plateau (see Figs.~4,5) and the multiplicity
rise, but most sensitive are correlations in
the relative polar angle
$\vth_{12}$ which are enhanced and suppressed
at small and large relative angles.
\par\vspace{0.4cm}
\noindent
{\sl 4. Running $\alpha_s$}. The distributions in various observables
are markedly different for a fixed or running $\as$ calculation, as
in case of momentum spectra and in particular
angular observables which are
power behaved for fixed $\as$.
 
In this review we have considered results from $\epem$ annihilations
but interesting results on color coherence phenomena are coming now also from
the hadron collider at Fermilab \cite{CDF} and at this conference
from HERA \cite{HERA}.
 
Analytical QCD calculations on multiparticle final states can reveal
interesting scaling and universality patterns
which provide new insights into the
intrinsic structures of jets. With sufficient care in the selection
of infrared safe quantities there is a promising path
in the calculation of multiparticle phenomena from basic principles
of QCD. Also the extension towards small scales $Q_0\sim m_h$ is worth
pursuing. A statisfactory understanding though of why PHD or LPHD
works so well is not really known.\par
\vspace{0.6cm}
\noindent{\bf 9. References}
\par%\vspace{0.4cm}

\end{document}